\documentclass[11pt]{article}
\usepackage{iap2000,epsfig}

\def\Journal#1#2#3#4{{\em #1} {\bf #2}, #3 (#4).}

\def\Note#1#2{#1 (#2).}

\def\be{\begin{equation}}
\def\ee{\end{equation}}
\def\bea{\begin{eqnarray}}
\def\eea{\end{eqnarray}}

\begin{document}

\vspace*{4cm}

\title{CONSTRAINING THE COSMOLOGICAL PARAMETERS FROM GRAVITATIONAL LENSES WITH SEVERAL FAMILIES OF IMAGES}

\author{ G. GOLSE, J.-P. KNEIB and G. SOUCAIL }

\address{Laboratoire d'Astrophysique, UMR 5572, Observatoire Midi-Pyr\'en\'ees\\ 
14 avenue E.-Belin, F-31400 Toulouse, France}

\maketitle\abstracts
{ The knowledge of the redshift of multiple images in cluster-lenses
allows to determine precisely the total projected mass within the
Einstein radius. The observation of various multiple images in a same
cluster is opening new possibilities to constrain the curvature of the
universe. Indeed, although the influence of $\Omega_m$ and
$\Omega_\lambda$ on the images formation is of the second order,
observations of many multiple images at different redshifts formed by
a regular cluster-lens should allow to constrain very accurately the
mass distribution of the cluster and to start to be sensitive to the
cosmological parameters entering the diameter angular distances.  We
present, analytical expressions and numerical simulations that
allow us to compute the expected error bars on the cosmological
parameters provided an HST/WFPC2 resolution image and spectroscopic
redshifts for the multiple images. Numerical tests on simulated data
confirm the rather small uncertainties we could obtain this way for the
two popular cosmological world models: $\Omega_m=0.3{\pm 0.24}$,
$\Omega_\lambda=0.7{\pm 0.5}$ or $\Omega_m=1.{\pm 0.33}$,
$\Omega_\lambda=0.{\pm 1.2}$. Our method can be applied from now on, on real clusters.}

\section{Introduction}

Recent works on constraining the cosmological parameters using the CMB
and the high redshift supernovae seem to converge to a new  ``standard
cosmological model'' favouring a flat universe with $\Omega_m\sim 0.3$
and $\Omega_\lambda\sim 0.7$: White~\cite{White} and references therein.
 However these
results are still uncertain and depend on some physical assumptions, so 
the flat $\Omega_m=1$ model is still
possible (Le Dour {\it et al.}~\cite{LeDour}). It is therefore
important to explore other independent techniques to constrain these
cosmological parameters.

In cluster gravitational lensing, the existence of multiple images -- with known redshifts -- given by
the same source allows to calibrate in an absolute way the total
cluster mass deduced from the lens model. The great improvement in the
mass modeling of cluster-lenses that includes the cluster galaxies
halos (Kneib {\it et al.}~\cite{Kneib96}, Natarajan \& Kneib~\cite{Natarajan})
 leads to the
hope that clusters can also be used to constrain the geometry of
the Universe, through the ratio of angular size distances, which only
depends on the redshifts of the lens and the sources, and on the
cosmological parameters. The observations of cluster-lenses containing
large number of multiple images lead Link \& Pierce~\cite{Link}
(hereafter LP98) to investigate this expectation. They considered a
simple cluster potential and on-axis sources, so that images appear as
Einstein rings. The ratio of such rings is then independent of the
cluster potential and depends only on $\Omega_m$ and $\Omega_\lambda$,
assuming known redshifts for the sources. According to them, this
would allow marginal discrimination between extreme cosmological
cases. But real gravitational lens systems are more complex concerning
not only the potential but also off-axis positions of sources. They
conclude that this method is ill-suited for application to real
systems.

We have re-analyzed this problem building up on the modeling
technique developed by us. As demonstrated below, we reach a rather different
conclusion showing that it is possible to constrain $\Omega_m$ and $\Omega_\lambda$ using the
positions of multiple images at different redshifts and some physically motivated lens models.

Troughout this paper we have assumed $H_0=65$ km s$^{-1}$ Mpc$^{-1}$, however the proposed method is independant of the value of $H_0$.

\section{Influence of $\Omega_m$ and $\Omega_\lambda$ on the images formation}

\subsection{Angular size distances ratio term}

\begin{figure*}[h]
\psfig{figure=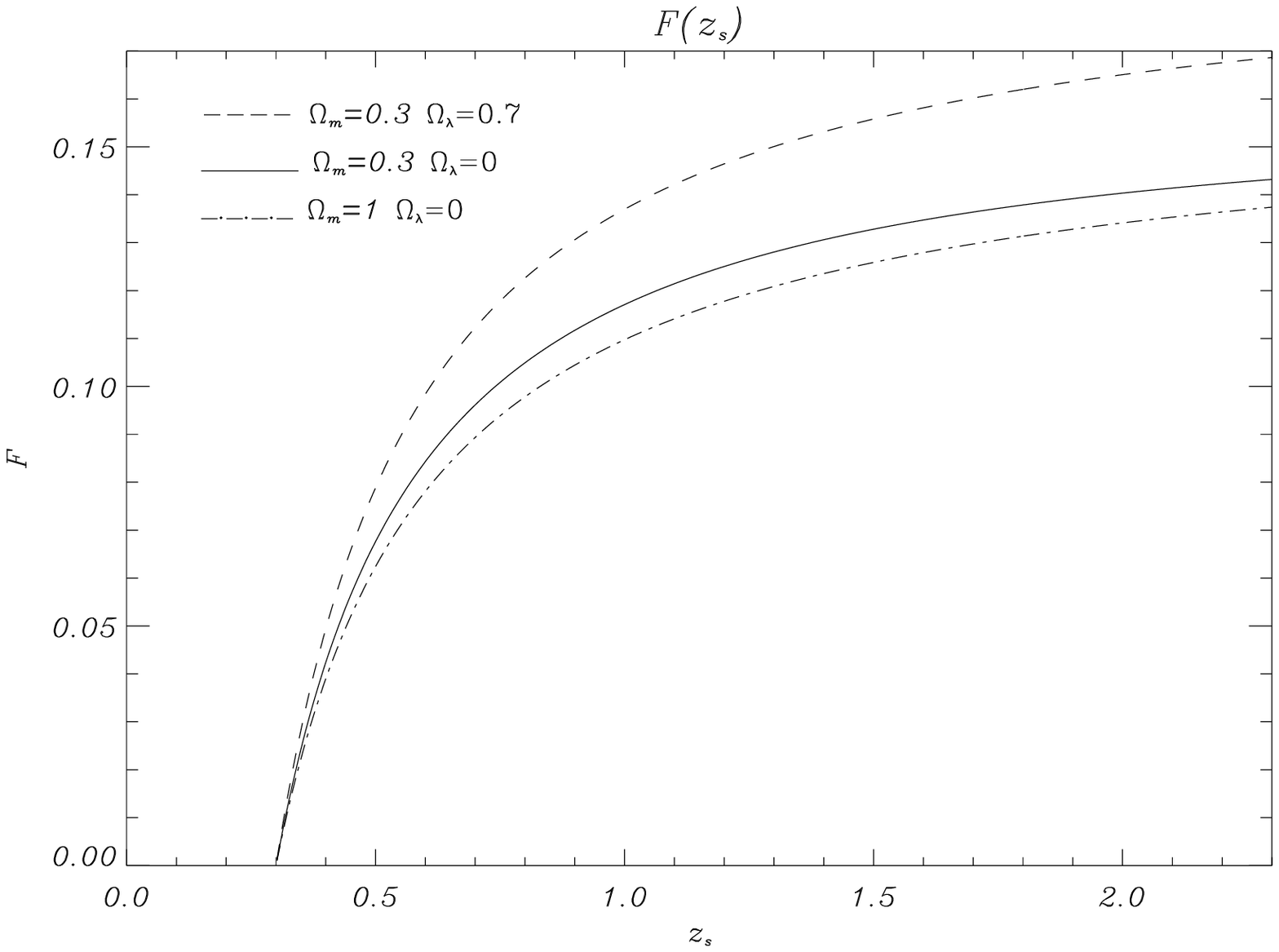, height=2.3in}
\hfill
\psfig{figure=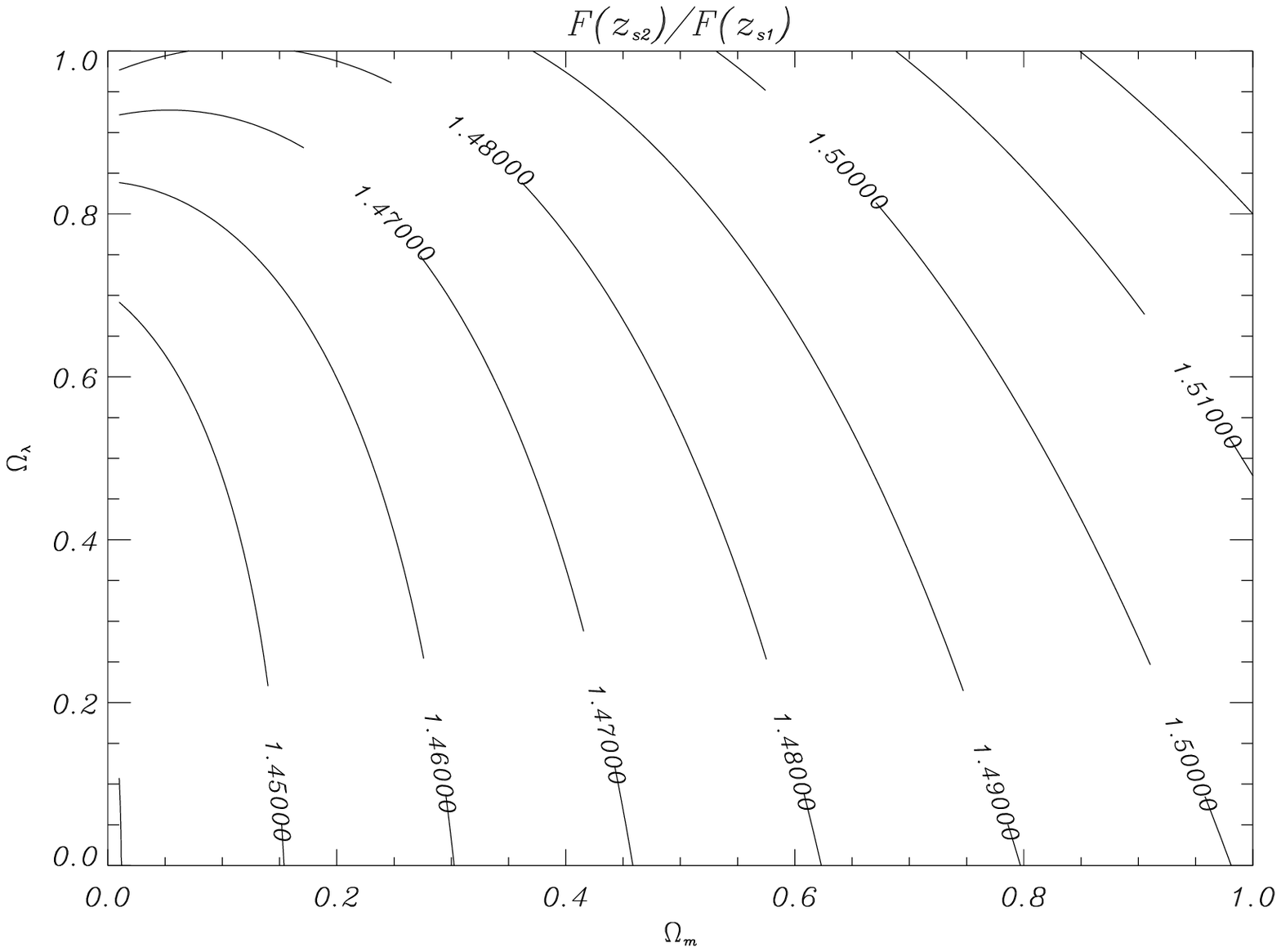, height=2.3in}
\caption{
\label{F_zs}
  Left. $F(z_s)$ for $z_l=0.3$ and various cosmological models.
  Right. $F(z_{s2})/F(z_{s1})$ as a function of $\Omega_m$ and
  $\Omega_\lambda$ for $z_l=0.3$, $z_{s1}=0.7$ and $z_{s2}=2$.  
}
\end{figure*}

In the lens equation: $\mathbf{\theta_{S}}= \mathbf{\theta_{I}} -
\displaystyle{\frac{2}{c^2}\frac{D_{OL}D_{LS}}{D_{OS}}} \mathbf\nabla 
\phi_\theta(\mathbf{\theta_{I}}) $, the dependence on $\Omega_m$ and
$\Omega_\lambda$ is solely contained in the term
$F=\displaystyle{{D_{OL}}{D_{LS}}/{D_{OS}}}$.
For a given lens plane, $F(z_s)$ increases rapidly up to a certain
redshift and then stalls, with significant differences for various
values of the cosmological parameters (see Fig. \ref{F_zs}).
Thus in order to constrain the actual shape of $F(z_s)$ several
families of multiple images are needed, ideally with their redshifts
regularly distributed in $F(z_s)$ to maximize the range in the $F$
variation.

If we consider fixed redshifts for both the lens and the sources, at least 2
multiple images are needed to derive cosmological constraints. In that case
$F$ has only an influence on the modulus of
$\mathbf{\theta_{I}}-\mathbf{\theta_{S}}$. So taking the ratio of two
different $F$ terms provides the intrinsic dependence on cosmological
scenarios, independently of $H_0$.
A typical configuration leads to the Fig. \ref{F_zs} plot. The
discrepancy between the different cosmological parameters is not very
large, less than 3\% between an EdS model and a flat low matter
density one. The figure also illustrates the expected degeneracy of the method, also confirmed
by weak lensing analyzes, with a continuous distribution of
background sources ({\it e.g.} Lombardi \& Bertin~\cite{Lombardi} ).

\subsection{Relative influence of the different parameters}

We now look at the relative influence of the different
parameters, including the lens parameters, to derive expected error bars
on $\Omega_m$ and $\Omega_\lambda$. To model the potential we choose
the mass density distribution proposed by Hjorth \& Kneib~\cite{Hjorth}, 
characterized by a core radius, $a$, and a cut-off radius
$s\gg a$. We can then get the expression of the deviation angle
modulus
$D_{\theta_{I}}=\parallel\mathbf{\theta_{I}}-\mathbf{\theta_{S}}\parallel$.

For 2 families of multiple images, the relevant quantity becomes the
ratio of 2 deviation angles for 2 images $\theta_{I1}$ and
$\theta_{I2}$ belonging to 2 different families at redshifts $z_{s1}$
and $z_{s2}$. Let's define
$R_{\theta_{I1},\theta_{I2}}=\displaystyle{\frac{D_{\theta_{I1}}}{D_{\theta_{I2}}}}$. With
several families, the problem is highly constrained because a single
potential must reproduce the whole set of images. In practice we
calculate
$\displaystyle{\frac{dR_{\theta_{I1},\theta_{I2}}}{R_{\theta_{I1},\theta_{I2}}}}$
versus the different parameters it depends on. We chose a typical
configuration to get a numerical evaluation of the errors on the cosmological parameters: $z_l=0.3$, $z_{s1}=0.7$,
$z_{s2}=2$, $\displaystyle{\frac{\theta_{I2}}{\theta_{I1}}}=2$,
$\displaystyle{\frac{\theta_{s}}{\theta_{a}}}=10$
($\theta_a=a/D_{OL}$,$\theta_s=s/D_{OL}$) and we assume $\Omega_m=0.3$
and $\Omega_\lambda=0.7$. We then obtain the following orders of magnitudes
for the different contributions :

\be
\displaystyle{\frac{dR_{\theta_{I1},\theta_{I2}}}{R_{\theta_{I1},\theta_{I2}}}} = 0.57\displaystyle{\frac{dz_{l}}{z_{l}}} + 0.74\displaystyle{\frac{dz_{s1}}{z_{s1}}} + 0.17\displaystyle{\frac{dz_{s2}}{z_{s2}}}
+ 0.4\left(\displaystyle{\frac{d\theta_{I1}}{\theta_{I1}}} - \displaystyle{\frac{d\theta_{I2}}{\theta_{I2}}}\right)
- 0.1\displaystyle{\frac{d\theta_{a}}{\theta_{a}}}
- 0.06\displaystyle{\frac{d\theta_{s}}{\theta_{s}}}
- 0.015\displaystyle{\frac{d\Omega_{m}}{\Omega_{m}}}
 + 0.02 \displaystyle{\frac{d\Omega_{\lambda}}{\Omega_{\lambda}}}
\ee

As expected, even with 2 families of multiple images the influence of
the cosmological parameters is of the second order. The precise value
of the redshifts is quite fundamental, therefore a spectroscopic
determination ($dz=0.001$) is essential. The position of the
(flux-weighted) centers of the images are also important. With HST
observations we assume $d\theta_I=0.1$''.

\begin{figure*}[h]
\psfig{figure=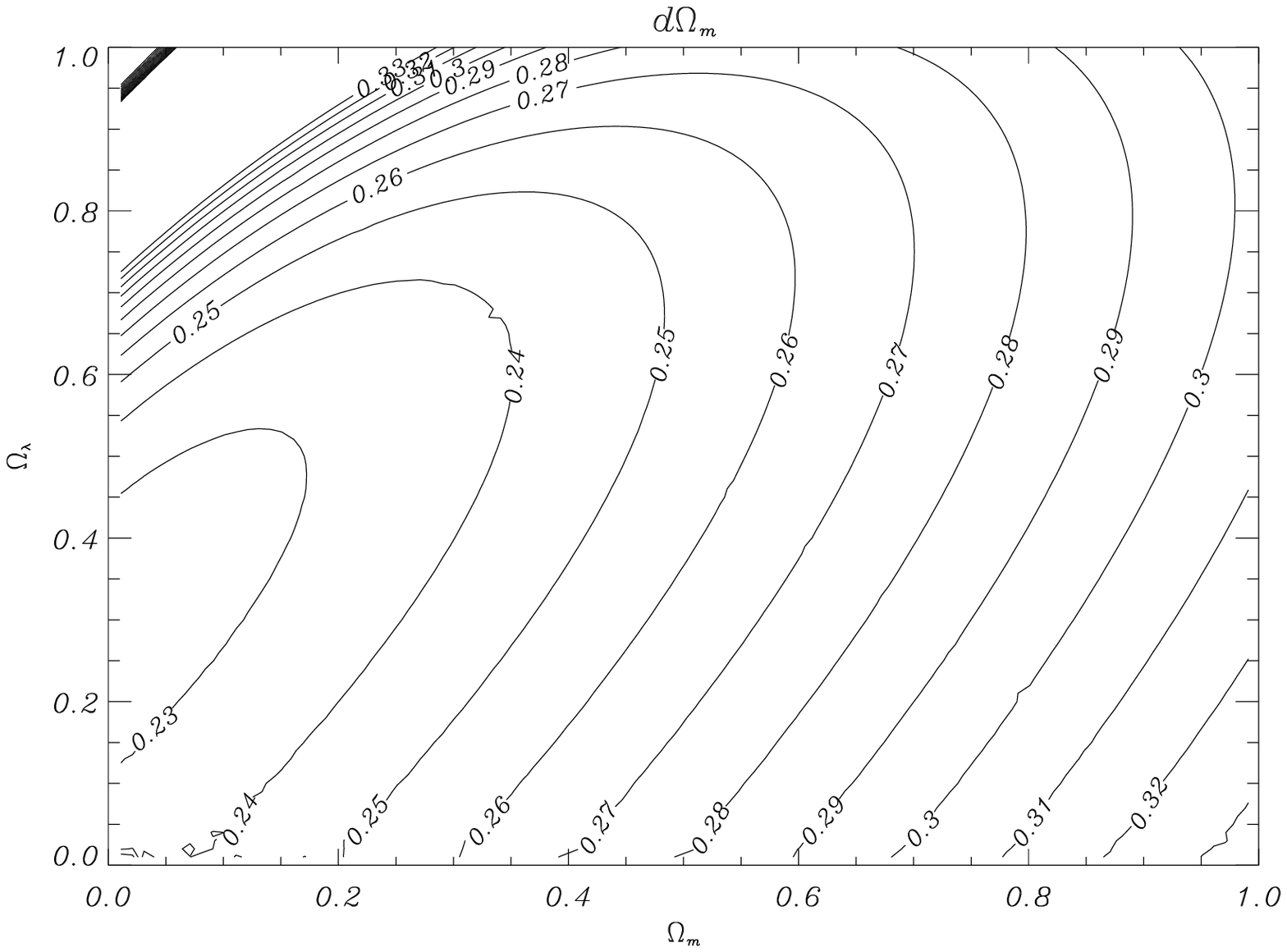, height=2.3in}
\hfill
\psfig{figure=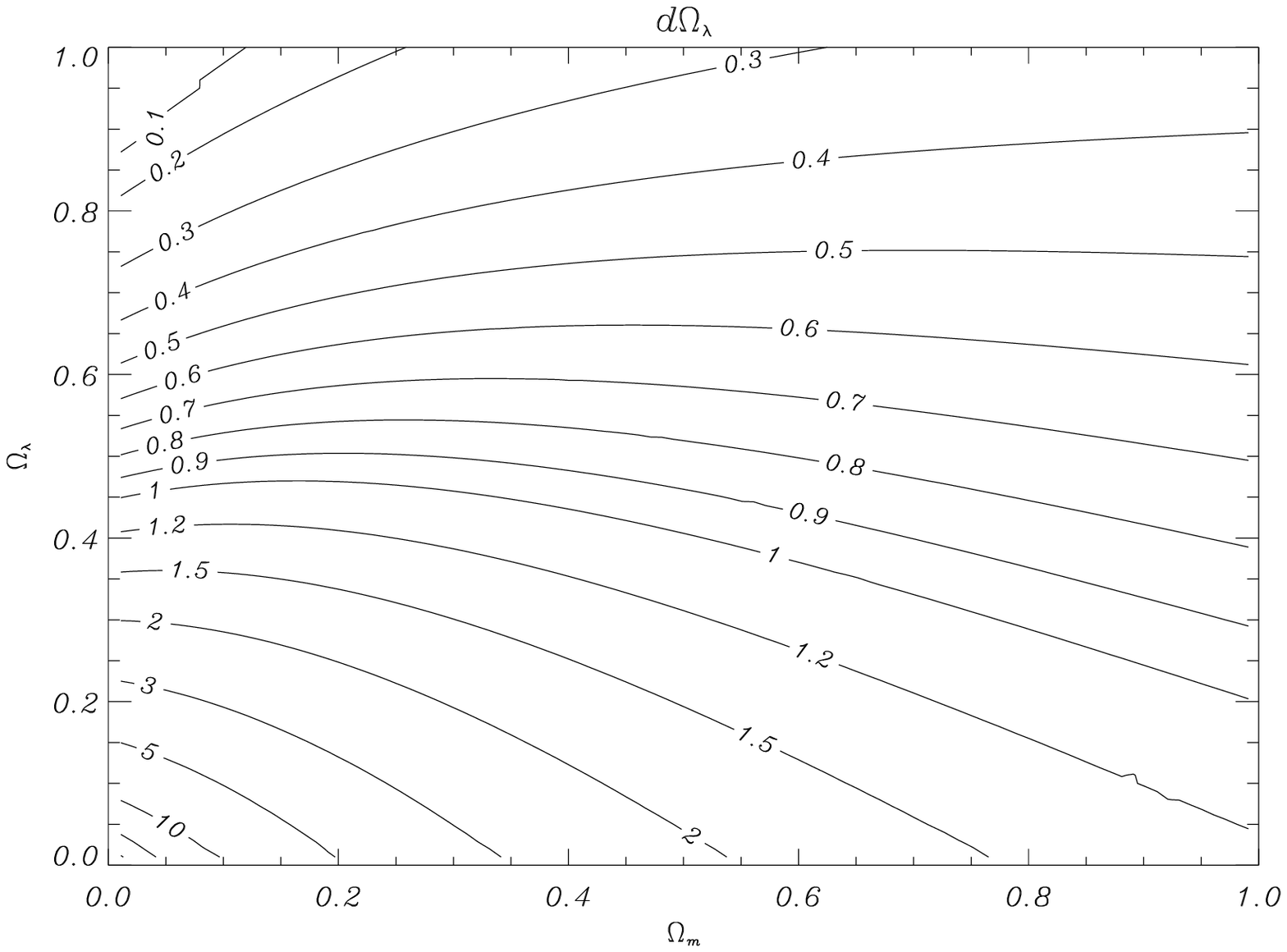, height=2.3in}
\caption{
\label{erreurs}
Expected error bars $d\Omega_m$ (Left) and $d\Omega_{\lambda}$ (Right) depending on the cosmological scenario with this method and in a typical case (see text).
}
\end{figure*}

So even if the problem is less dependent on the core and cut-off
radii (in other word the mass profile), they will represent the main sources of error. Taking
$d\theta_a/\theta_a= d\theta_s/\theta_s= 20$ \%, we then derive
the errors $d\Omega_m$ and $d\Omega_{\lambda}$ from the above relation
in the flat low matter density we chose. We did this computation for different sets of cosmological models. Indeed the errors we will obtain with this method change significantly with respect to $\Omega_m$ and $\Omega_\lambda$. All other things being equal apart from the cosmological parameters, we plot $d\Omega_m$ and $d\Omega_\lambda$ for a continuous set of universe models (Fig. \ref{erreurs}). For instance in the 2 popular cosmological scenarios, we have :

$\Omega_m=0.3\pm0.24 $ \hspace{0.5cm} $\Omega_\lambda=0.7\pm0.5$
\hspace{1cm} or \hspace{1cm} $\Omega_m=1\pm0.33$ \hspace{0.5cm}
$\Omega_\lambda=0\pm1.2$

As this can be easily understood from the Fig. \ref{F_zs} degeneracy plot, the method is in general far more sensitive to the matter density than to the cosmological constant, for which the error bars are larger.

However the results we could obtain this way are as precise as the ones given by other constraints. But these errors are just typical; provided spectroscopic and HST observations, they depend mostly on the particular cluster and the potential model chosen to describe it. They could be quite tightened with a precise model, and by increasing the number of clusters with multiple images.

\section{\label{simul}Constraint on $(\Omega_m,\Omega_\lambda)$ from strong lensing}

\subsection{Method and algorithm for numerical simulations }

We consider basically the potential introduced in section 2.2. After considering the lens equation, fixing arbitrary
values $(\Omega_m^0$,$\Omega_\lambda^0)$ and a cluster lens redshift
$z_l$, our code can determine the images of a
source galaxy at a redshift $z_s$.
Then taking as single observables these sets of images as well as the
different redshifts, we can recover some parameters (the more
important ones being $\sigma_0$, $\theta_a$ or $\theta_s$) of the
potential we left free for each point of a grid
$(\Omega_m$,$\Omega_\lambda)$. The likelihood of the result is
obtained via a $\chi^2$-minimization, where the $\chi^2$ is computed
in the source plane as follows :

\be
\chi^2=\displaystyle{\sum_{i=1}^n \sum_{j=1}^{n^i} \frac{(\theta_{Sj}^i-\theta_{SG}^i)^2}{\sigma_{Si}^{j\,2}}}
\ee

 The subscript $i$ refers to the families and the subscript $j$ to the images of a family. There is a total of $\sum_{i=1}^n n^i=N$ images. $\theta_{Sj}^i$ is the source found for the image $\theta_{Ij}^i$ in the inversion. $\theta_{SG}^i$ is the barycenter of the $\theta_{Sj}^i$ (belonging to a same family). Finally if $\sigma_{Ii}^{j}$ is the error on the position of the center of  $\theta_{Ij}^i$ and $A_j^i$ the amplification for this image, then $\sigma_{Si}^{j}=\sigma_{Ii}^{j}/\sqrt{A_j^i}$.

\subsection{Numerical simulations in a typical configuration}

To recover the parameters of the potential ({\it i.e.}
$\sigma_0$, $\theta_a$, $\theta_s$ and adjusted lens parameters), we generated
 3 families of
images with regularly distributed source redshifts.

\begin{figure*}
\psfig{figure=Image7.ps, height=2.4in}
\hfill
\psfig{figure=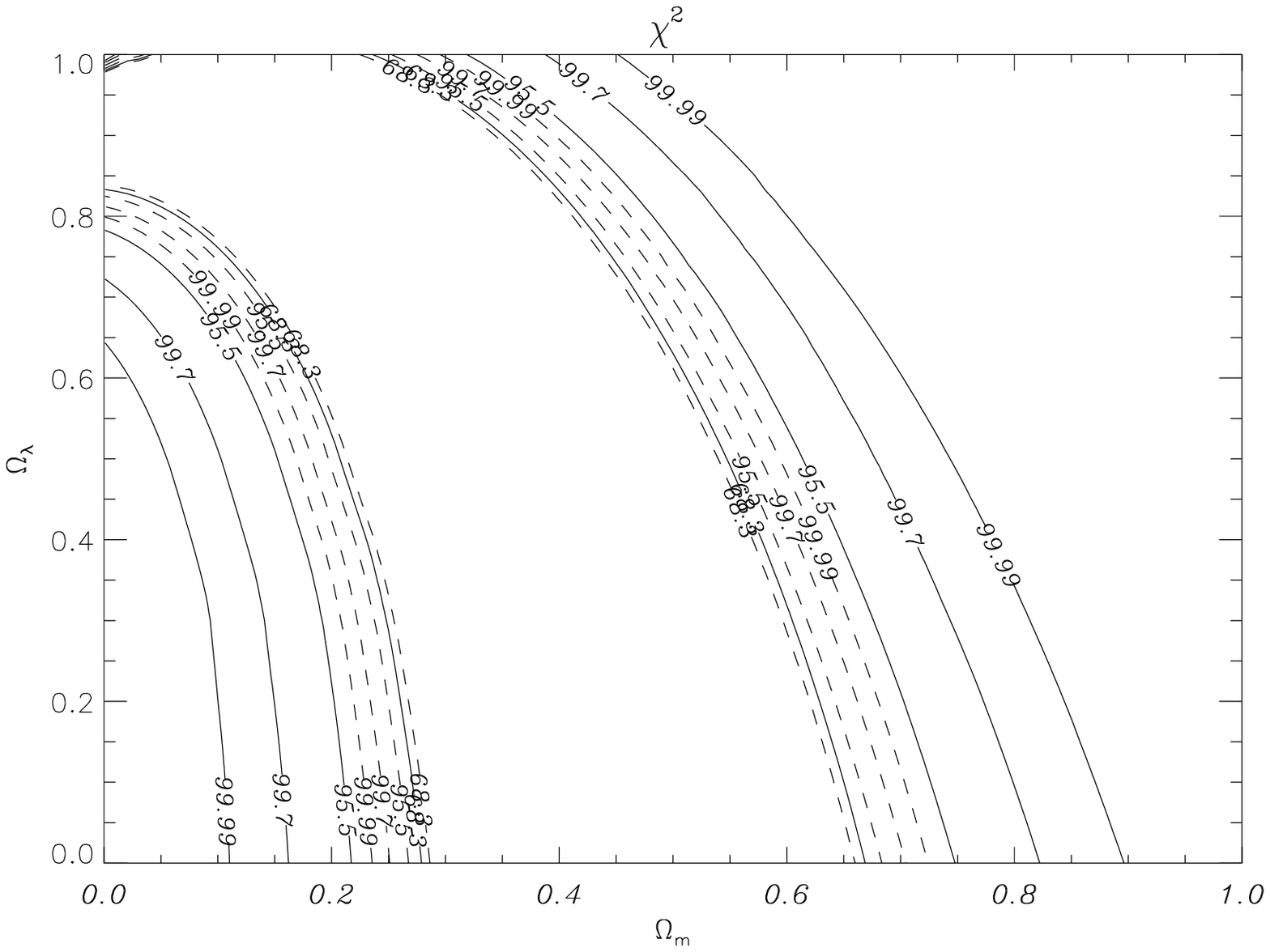, height=2.4in}
\caption{\label{3fam}
  {Left. Generation of images by a $z_l=0.3$ cluster with
  $\sigma_0=1400$ km/s, $\theta_a=13.54$'' and
  $\theta_s=145.8$''. Close to their respective critic lines, we see 3
  families of images at $z_{s1}=0.6$, $z_{s2}=1$ and $z_{s3}=2$.
  Right. Solid lines : $\chi^2(\Omega_m,\Omega_\lambda)$ confidence
  levels obtained for this configuration. Generating arbitrary values:
  $(\Omega_m^0,\Omega_\lambda^0)=(0.3,0.7)$. Dashed lines:
  $\chi^2(\Omega_m,\Omega_\lambda)$ confidence levels obtained
  considering 10 clusters in this same configuration.}}
\end{figure*}

For starting values
$(\Omega_m^0,\Omega_\lambda^0)=(0.3,0.7)$ we obtained
confidence levels shown in Fig. \ref{3fam}. The method puts forward a good
constraint, better on $\Omega_m$ than on $\Omega_\lambda$, and the
degeneracy is the expected one (see Fig. \ref{F_zs}). Concerning the
free parameters, we also recovered in a rather good way the potential,
the variations being $\Delta\sigma_0\sim150$ km/s,
$\Delta\theta_a\sim3$''and $\Delta\theta_s\sim20$''.

\begin{figure*}
\hspace{3cm}
\psfig{figure=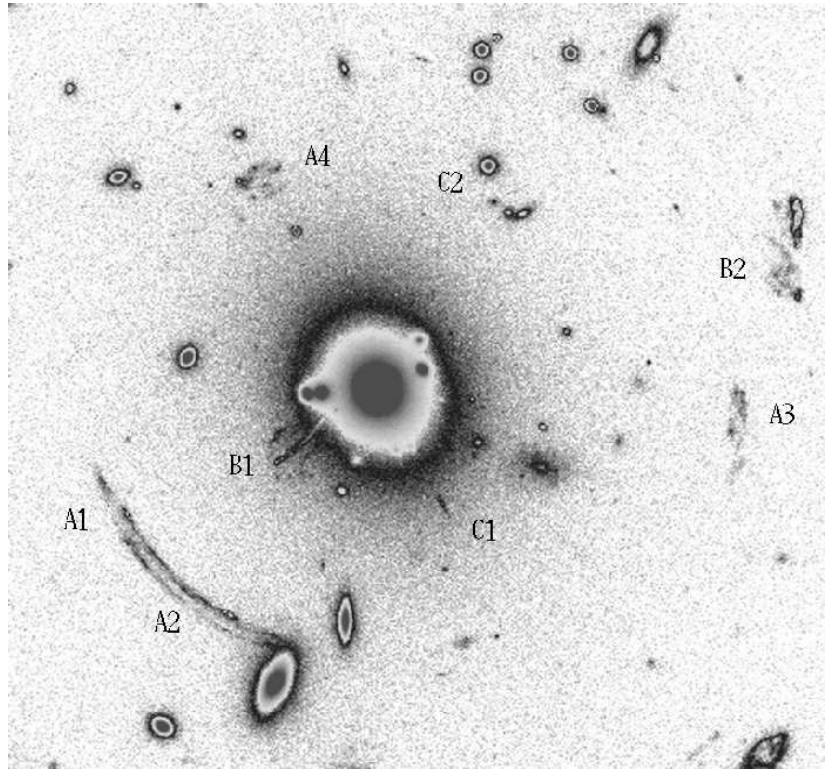, height=4in}
\caption{\label{MS2137}
  {HST image of MS2137-23 ($z_l=0.313$). We see 3 families of images, a tangential arc A1-A2 with its associated arcs A3 and A4, and 2 radial arcs B1 and C1 with their associated images B2 and C2.
}}
\end{figure*}

This is an ``ideal'' case, of course, because we tried to recover the
same type of potential we used to generate the images, the morphology
of the cluster being quite regular and the redshift range of the
sources being wide enough to check each part of the
$F$ curve.
Such a simple approach can be applied to regular clusters like
MS2137-23, which shows at least 3 families of multiple images
including a radial one, see Fig. \ref{MS2137}. But the spectroscopic redshifts are still currently missing for this cluster.

\section{Conclusions \& prospects}

Following the work of LP98, we discussed a method to obtain
informations on the cosmological parameters $\Omega_m$ and
$\Omega_\lambda$ while reconstructing the lens gravitational potential
of clusters with multiple image systems at different redshifts.

This technique gives degenerate constraints, $\Omega_m$ and
$\Omega_\lambda$ being negatively correlated, with a better constraint
of the matter density. With a single cluster in a typical lensing
configuration we can expect the following error bars :
$\Omega_m=0.3{\pm 0.24}$, $\Omega_\lambda=0.7{\pm 0.5}$. To perform
that, several general conditions must be fulfilled: 

$\ast$ a cluster with a rather regular morphology, so that a few parameters are needed to describe the gravitational potential ; this is not so restrictive because we saw that a bimodal cluster can also provide a constraint,

$\ast$ ``numerous'' systems of multiple images, probing each part of the cluster,

$\ast$ a good spatial resolution image (HST observations) of the cluster and arcs to compute relatively precise -- 0.1'' -- (flux weighted) images positions,

$\ast$ spectroscopic precision for the different redshifts that should be also regularly distributed from $z_l$ to high values -- this requires deep spectroscopy on 8-10m class telescopes due to the faintness of the multiple images .

It is important to notice that one cluster could provide one constraint on the geometry of the whole universe. And it is possible to combine data from different clusters to tighten the error bars. Combining the study of about 10 clusters would lead to meaningful constraints. The dashed lines confidence levels in the Fig. \ref{3fam} are the result of a numerical simulation made with 10  {\it identical} clusters.

  Actually the degeneracy depends only on the different redshifts involved that we will have various sets of when applying the method to real configurations. This should lead to a more reduced area of allowed cosmological parameters. We are encouraged by more and more known observations including systems with multiple sources and we plan to apply this technique to clusters like MS2137-23, MS0440+02, A370, AC114 and A1689.

\end{document}